\documentclass[aps,prl,reprint,groupedaddress,showpacs]{revtex4-1}

\usepackage{graphicx,epstopdf}
\usepackage{amsmath}

\begin{document}


\title{A Ball Breaking Away from a Fluid}


\author{B.~Turnbull}
\affiliation{Faculty of Engineering, University of Nottingham, University Park, NG7 2RD. UK}
\author{M.~M.~Scase}
\author{D.~S.~Percival}
\affiliation{School of Mathematical Sciences, University of Nottingham, University Park, NG7 2RD. UK}


\date{\today}

\begin{abstract}
We consider the withdrawal of a ball from a fluid reservoir to understand the longevity of the connection between that ball and the fluid it breaks away from, at intermediate Reynolds numbers. Scaling arguments based on the processes observed as the ball interacts with the fluid surface were applied to the `pinch-off time', when the ball breaks its connection with the fluid from which it has been withdrawn, measured experimentally. At the lowest Reynolds numbers tested, pinch-off occurs in a `surface seal' close to the reservoir surface, where at larger Reynolds numbers pinch-off occurs in an `ejecta seal' close to the ball. Our scaling analysis shows that the connection between ball and fluid is controlled by the fluid film draining from the ball as it continues to be winched away from the fluid reservoir. The draining flow itself depends on the amount of fluid coating the ball on exit from the reservoir. We consider the possibilities that this coating was created through: a surface tension driven Landau Levitch Derjaguin wetting of the surface; a visco-inertial quick coating; or alternatively through the inertia of the fluid moving with the ball through the reservoir. We show that although the pinch-off mechanism is controlled by viscosity, the coating mechanism is governed by a different length and timescale, dictated by the inertial added mass of the ball when submersed. 
\end{abstract}


\maketitle

\begin{figure}[htbp]
\includegraphics[width=0.6\columnwidth]{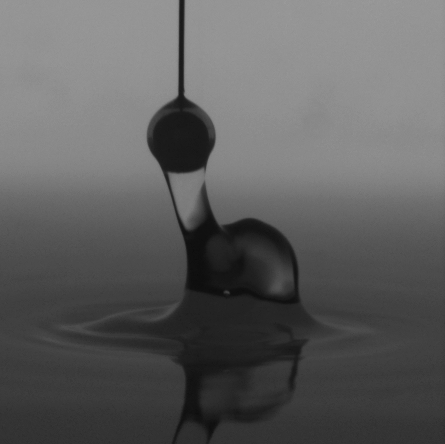}
\caption{A 6 \,mm diameter steel ballbearing winched from a water bath at 0.5\,m/s. The ball extrudes a tendril of fluid behind it, while the motion of the ball has also formed a surface structure alongside.}\label{fig:eject}
\end{figure}

The process of withdrawing an object from a liquid reservoir can be viewed as the converse process to a solid body entering a liquid reservoir (water-entry) \cite{enriquez2011non,truscott2014water}. As opposed to the solid drawing air into the fluid to create a cavity \cite{aristoff2009water}, fluid is extruded in the wake of the solid to form a connecting tendril, sometimes alongside other surface features (see e.g., Fig.~\ref{fig:eject}).  

Here, we consider intermediate Reynolds number flows ($\mathcal{O}(10^3)$), that may be considered incompressible. In this regime, the withdrawal of objects from reservoirs is common in manufacturing where a solid requires coating with a liquid layer that may solidify, such as when producing toffee apples or coating glass fibers \cite{townsend2015fluid,quere1999fluid,tallmadge1967entrainment}. 

The ejection of solids from liquids also occurs in nature. For example fish or dolphins leaping from water \cite{fish2006dynamics}, or in debris flows of rocks, water and fine sediment where the collisional behaviour of the rock flow is mediated by the sludgy fluid \cite{takahashi2014debris}. The capacity of the fluid to keep the rocks inside the flow is related to the longevity of the connection between a given rock and the fluid as that rock is collisionally ejected from the flow \cite{turnbull2015debris}. In such flows the presence of the fine sediments also leads to complex non-Newtonian bulk fluid rheology.  Our experiments begin to capture the effects of such intermediate Reynolds number flows, differing in behaviour from both viscous creeping flows and high Reynolds number cavitating flows.

\begin{figure}[htbp]
\includegraphics[width=0.6\columnwidth]{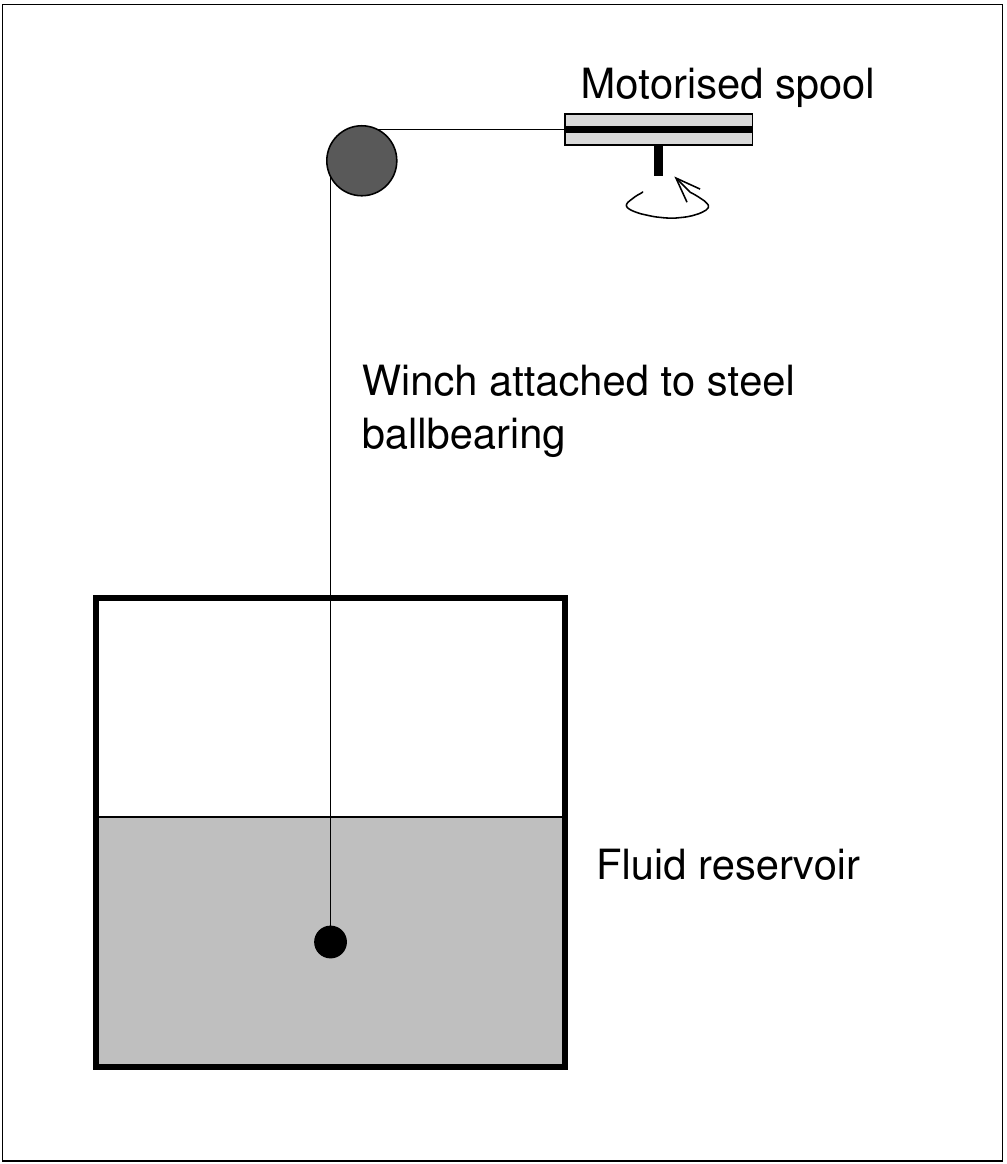}
\caption{Illustration of an experiment winching a ball from a fluid reservoir.}\label{fig:setup}
\end{figure}

In this letter we have analysed a series of experiments (illustrated in Fig.~\ref{fig:setup}) where a stainless steel ballbearing was winched from a reservoir of fluid, to understand which processes controlled the longevity of the connection between ball and fluid. The balls (diameters 3--12\,mm) were drilled through their centre with a 0.45\,mm hole into which a fine needle was fixed. That needle was threaded and the thread passed over a pulley to a spool. A motor spun the spool to give a range of speeds (0.2--0.6\,m/s) drawing the ball from the tank. The ball started from a hanging position supported by the winch and quickly accelerated to travel with constant speed well in advance of meeting the fluid surface ($>$ 4 ball diameters). Captured video sequences permitted the `pinch-off time' to be measured - that is the time from when the ball passed through the level of the fluid surface until the tendril connecting ball and fluid reservoir was broken.

\begin{figure*}[htbp]
\includegraphics[width=\textwidth]{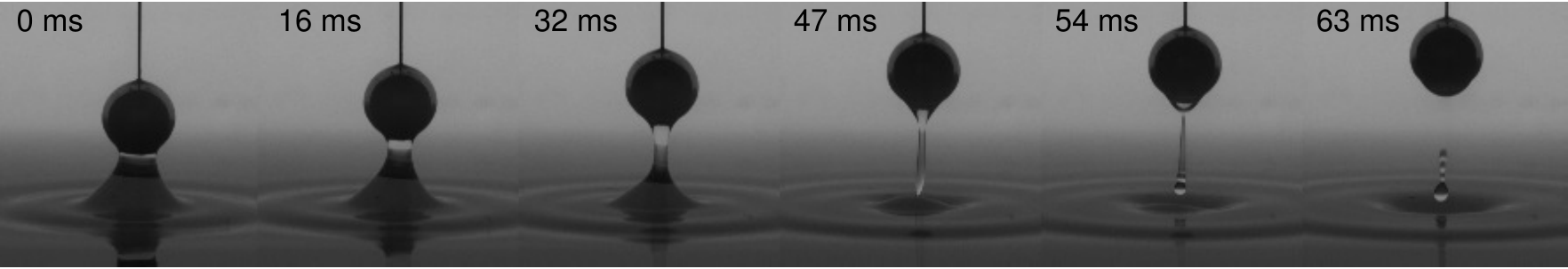}
\caption{A 12\,mm diameter ball winched from a water bath at 0.2\,m/s, time lapse image. The time from the first image is shown in milliseconds at the top of each panel. Note that the tendril connecting the ball to the fluid reservoir first pinches off in a surface seal, the converse of a shallow seal in the water-entry problem \cite{aristoff2009water}.}\label{fig:timelapse_slow}
\end{figure*}

\begin{figure*}[htbp]
\includegraphics[width=\textwidth]{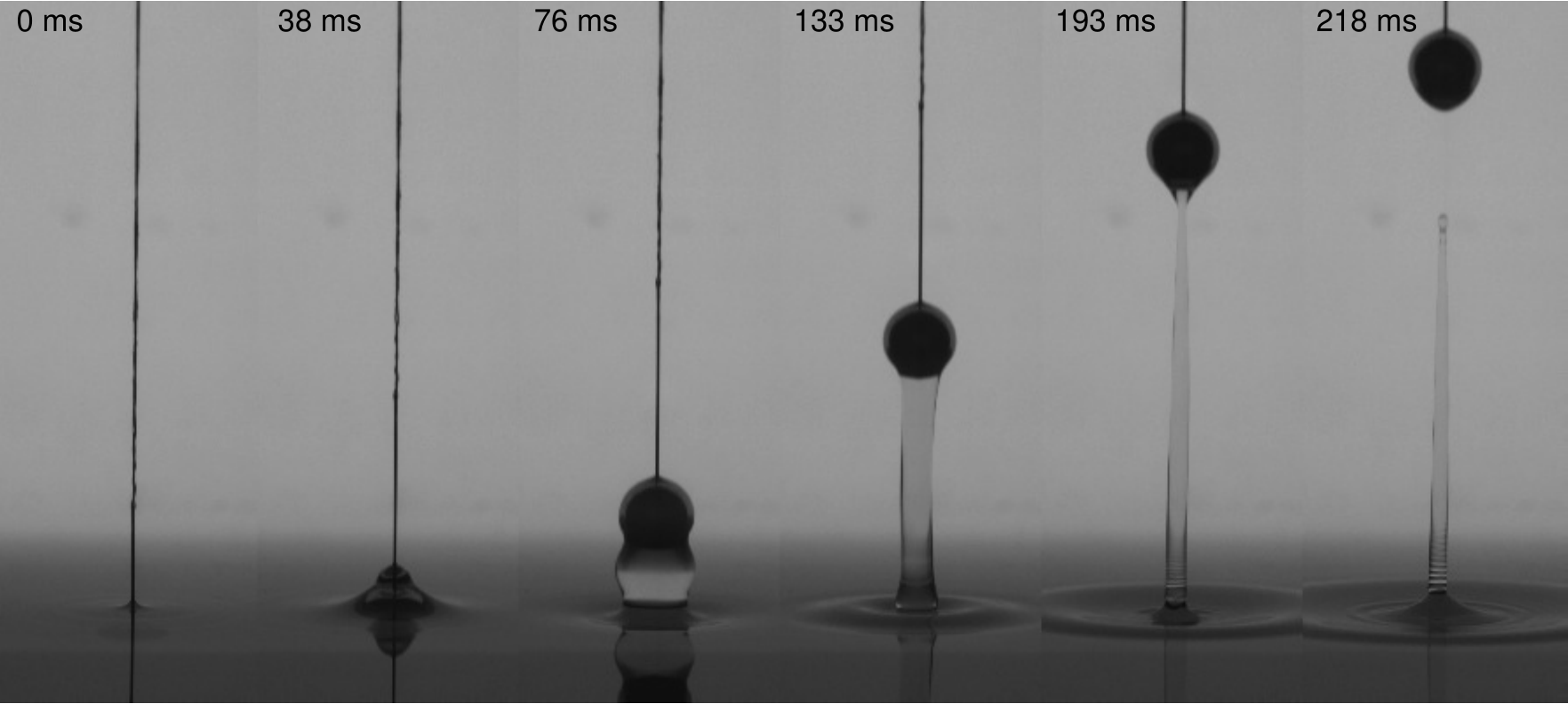}
\caption{ 12\,mm diameter ball winched from a water bath at 0.6\,m/s, time lapse image. The time from the first image is shown in milliseconds at the top of each panel. Bands of capillary waves at the base of the tendril \cite{hancock2002fluid} before it pinches off indicate that the fluid inside the tendril is flowing towards the water bath at this time. Pinch-off at the ball in an ejecta seal is the converse of a deep seal in the water-entry problem \cite{aristoff2009water}.}\label{fig:timelapse_fast}
\end{figure*}

\begin{figure*}[htbp]
\includegraphics[width=\textwidth]{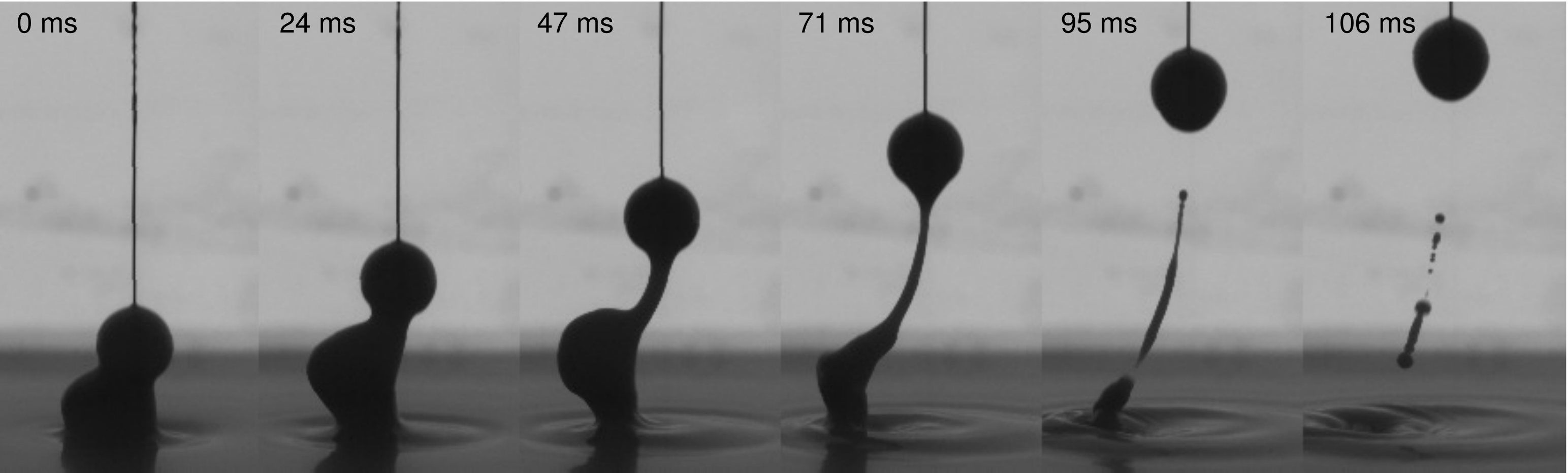}
\caption{ 12\,mm diameter ball winched from a 20\% kaolin suspension bath at 0.5\,m/s, time lapse image. The time from the first image is shown in milliseconds at the top of each panel. The tendril pinches off in an ejecta seal, with Rayleigh-Plateau instability seen as the tendril breaks into droplets \cite{eggers2008physics}.}\label{fig:timelapse_kaolin}
\end{figure*}

To investigate the effects of surface tension, fluid rheology and the presence of fine sediments \cite{ghosh2007spontaneous} on the evolving flow, the experiment was repeated using different fluids: water, glycerol solution (increasing the viscosity), a surfactant solution (reducing surface tension), and dilute non-Newtonian suspensions of kaolin powder and of custard powder. Each combination of ball size, fluid and winch speed was run at least 3 times with all of these repeated data points shown here, as an indicator of experimental error. Figs.~\ref{fig:timelapse_slow}, \ref{fig:timelapse_fast} and \ref{fig:timelapse_kaolin} show stills from videos of the 12\,mm diameter ball being winched from water at 0.2\,m/s and 0.6\,m/s and from a 20\% kaolin suspension at 0.5\,m/s respectively. These sequences show that the tendril grows longer with faster winch speeds and increasing ball size. Features reminiscent of the water-entry problem also manifest. The tendril can pinch-off close to the water surface in a `surface seal' - the converse of `shallow seal' in water-entry, or close to the ball in an `ejecta seal' -  the converse of `deep seal' in water-entry \cite{aristoff2009water}. The experiments with fine sediments in suspension (20\% kaolinite) show more complex surface features than the water experiments, with both surface and ejecta seals and also a Rayleigh-Plateau type break up of the tendril. 


\begin{figure*}
a)~\includegraphics[width=0.9\columnwidth]{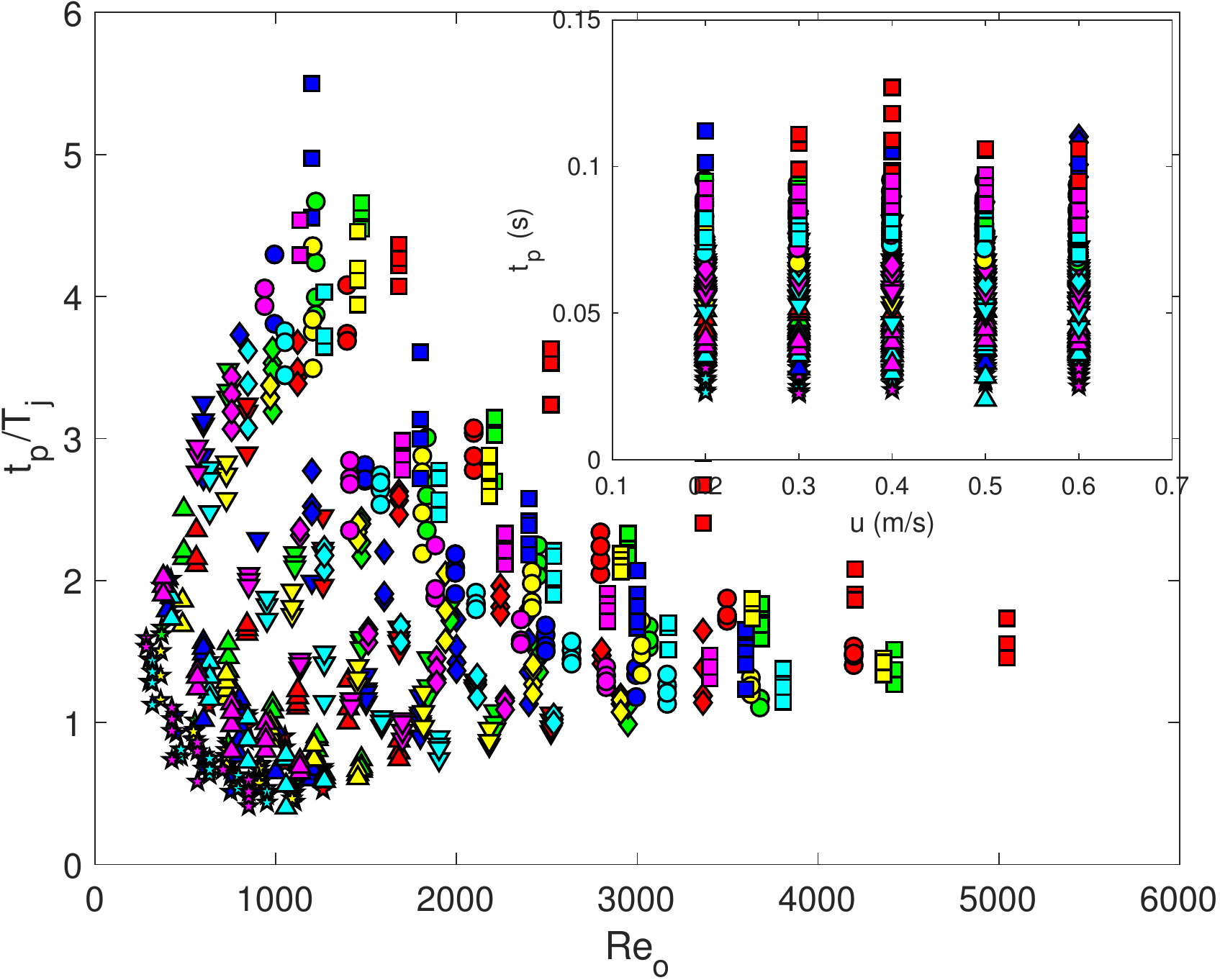}\quad b)~\includegraphics[width=0.9\columnwidth]{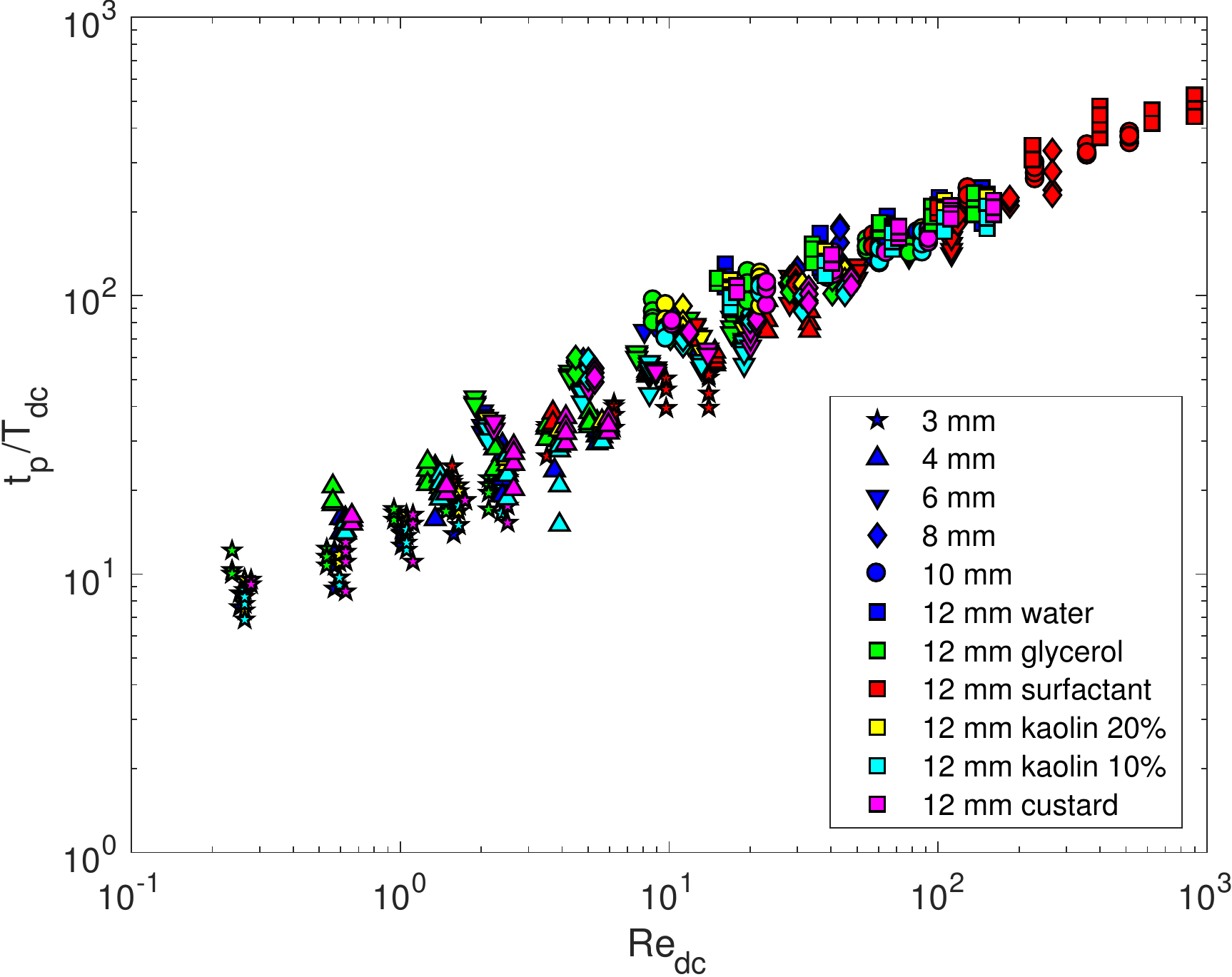}\\c)~\includegraphics[width=0.9\columnwidth]{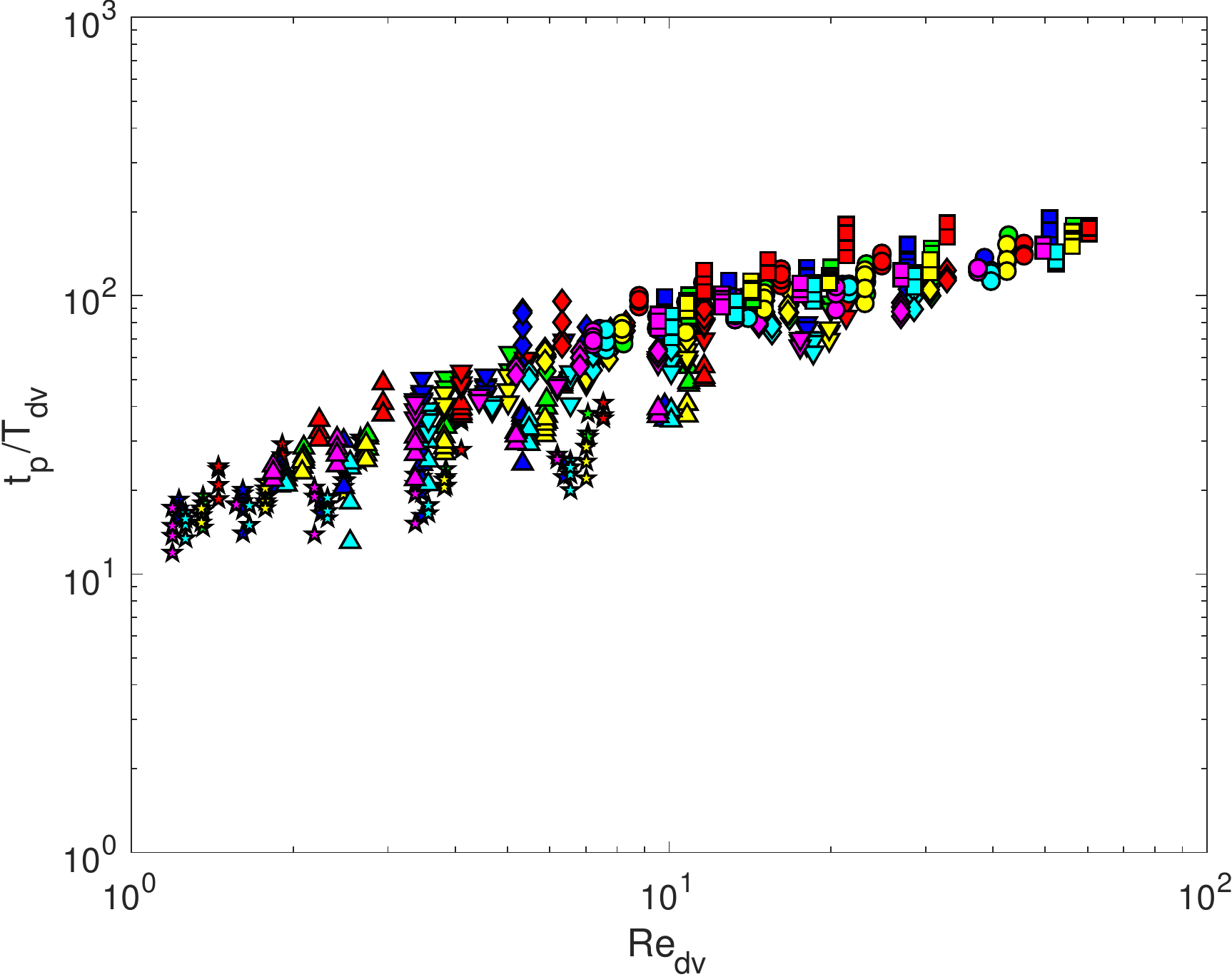}\quad d)~\includegraphics[width=0.9\columnwidth]{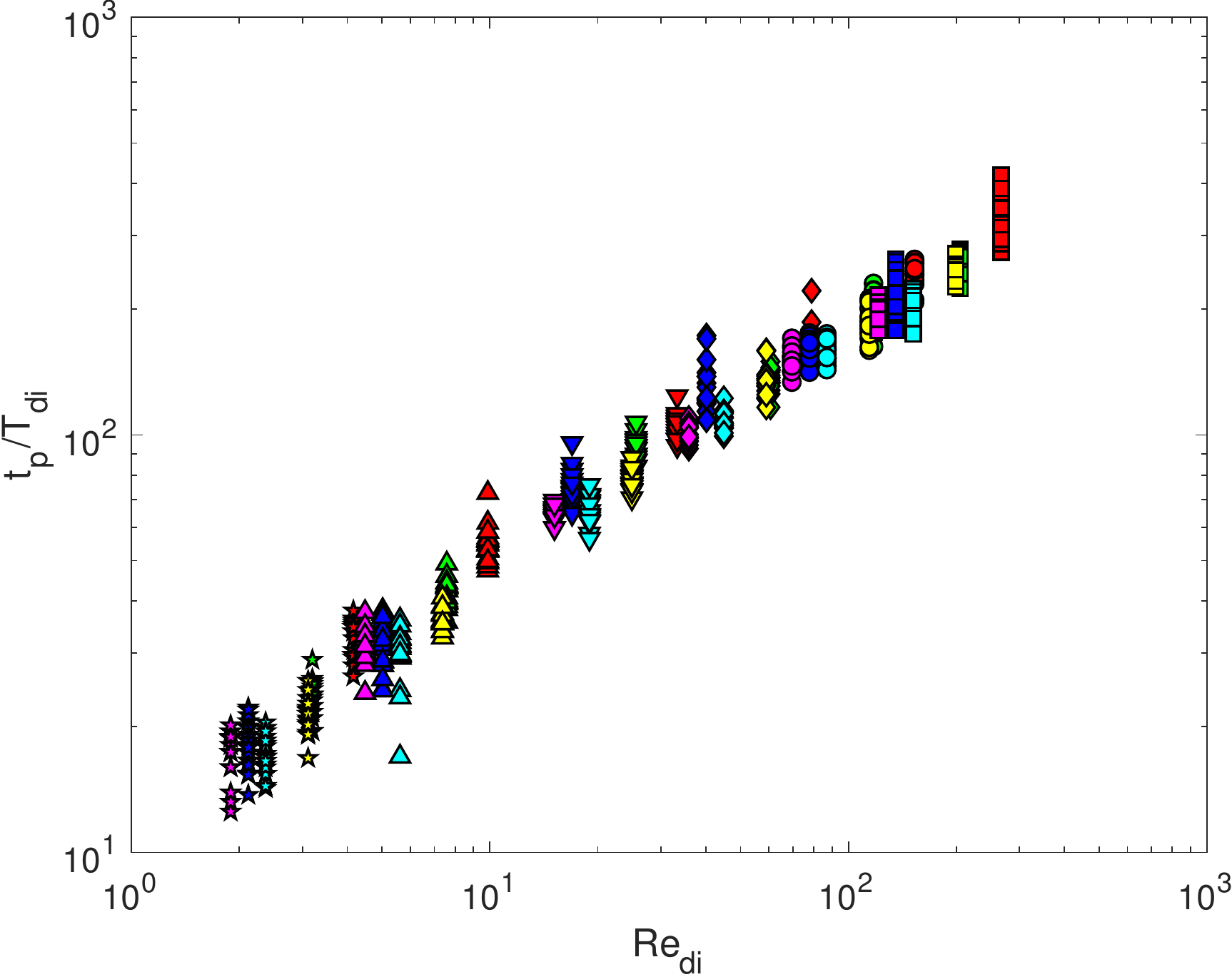}
\caption{a)~~Pinch-off time $t_p$ scaled with ejection jet time scale, $T_j=u/g$, versus Reynolds number ($\textnormal{Re}_o=uR/\nu$); inset, unscaled data. b)~Non-dimensional pinch-off time versus draining Reynolds number, Capillary coating scaling (Eqn.~\ref{eqn:dc}). c)~Non-dimensional pinch-off time versus draining Reynolds number, visco-inertial coating scaling (Eqn.~\ref{eqn:sca_vi}). d)~Non-dimensional pinch-off time versus draining Reynolds number, inviscid inertial coating scaling (Eqn.~\ref{eqn:di}).}\label{fig:scaling}
\end{figure*}

The observation that the winch sometimes goes slack (including in the formation of the structure seen in Fig.~\ref{fig:eject}) and PIV on the flow induced in the reservoir together indicate that a vortex ring in the wake of the ball becomes trapped by the fluid surface. Fluid is thus accelerated through the eye of this vortex ring, upwards. One possibility is that the longevity of the tendril is determined by this `ejection' flow. If this is the case, then the maximum height the ejection jet reaches follows from an energy balance with an associated timescale $T_j=u/g$, where $u$ is the winch speed and $g$ is the acceleration due to gravity. The pinch-off time non-dimensionalised by this timescale is shown in Fig~\ref{fig:scaling} a for the tested range of Reynolds numbers, $\textnormal{Re}_o = \rho u R/\mu$, where $R$, $\rho$ and $\mu$ are the ball radius, fluid density and dynamic viscosity respectively.

This ejection jet scaling does not collapse the data, indicating that an alternative mechanism is primarily responsible for extruding the tendril. 

As the ball is winched away from the fluid reservoir, it carries with it a fluid layer that drains into the top of the extending tendril. For the slowest winch speeds, the tendril pinches off in a surface seal (e.g.~at 54\,ms in Fig.~\ref{fig:timelapse_slow}) suggesting that the tendril is mostly filled from the draining flow rather than the ejection jet. At higher speeds (Fig.~\ref{fig:timelapse_fast}), bands of capillary waves at the base of the tendril just before pinch-off indicate that the fluid inside the tendril is flowing down towards the reservoir \cite{hancock2002fluid}, suggesting that any upward ejection of fluid is no longer important. 

If we assume gravitationally driven viscous Stokes flow in this very thin layer draining from the ball surface, the pressure evolves as
\begin{equation}
\nabla p=\rho\mathbf{g}+\mu\nabla^2\mathbf{u_d},
\end{equation}
where $\mathbf{u_d}$ is the fluid draining velocity, i.e.~the fluid velocity relative to the ball.  Thus, if the draining layer thickness $s$ is the characteristic length scale of the flow and $U_d$ a characteristic draining speed,
\begin{eqnarray}
\nonumber \rho g \sim \frac{\mu U_d}{s^2}
\quad \mathrm{implying} \quad U_d \sim \frac{\rho gs^2}{\mu}.
\end{eqnarray}
This leads to a draining layer Reynolds number with a corresponding draining timescale
\begin{equation}\label{eqn:draining}
\textrm{Re}_d \sim \frac{gs^3}{\nu^2}\quad \mathrm{and}\quad T_d = \frac{s}{U_d} \sim \frac{\nu}{gs},
\end{equation}
where $\nu=\mu/\rho$ is the fluid kinematic viscosity. Note that this draining Reynolds number may be very small, as required for Stokes flow, while the overall Reynolds number of the system, $\textnormal{Re}_o$, can be much larger because the film on the ball is much thinner than the ball radius and draining speeds much smaller than winch speeds.

To complete this model of the draining flow, we need to establish the characteristic thickness of the draining fluid layer, $s$, that starts as a coating deposited on the ball as it is winched through the reservoir surface level. We might expect this coating layer to follow a Landau, Levich \& Derjaguin model for the fluid layer on a solid plate or fiber \cite{white1965static,quere1999fluid,levich1942dragging,derjaguin1943thickness,de1985wetting} withdrawn from a bath. For a fluid with surface tension $\gamma$, the Capillary number is a ratio of viscous and surface tension forces, $\textnormal{Ca}=u\mu/\gamma$. As the ball passes through the reservoir surface level to become coated, the relevant velocity scale is the winch speed, leading to Capillary numbers $10^{-2}$ to $10^{-3}$ in our experiment. In this low Capillary number ($\textnormal{Ca}\ll1$) regime, surface tension dominates and the classical $s\propto R\textnormal{Ca}^\frac{2}{3}$ \cite{jin2005drag,quere1999fluid} may provide a characteristic length scale for the fluid layer. This law applies to plates at arbitrary inclination \cite{jin2005drag} and to long fibers, but will not capture the effects of a time-varying pressure gradient as the coating layer grows. The draining Reynolds number and timescale based on this capillary coating mechanism are thus
 \begin{equation}\label{eqn:dc}
\textnormal{Re}_{dc} \propto \frac{\textnormal{Ca}^2gR^3}{\nu^2} \quad \textrm{and} \quad T_{dc} \propto \frac{\nu}{Rg\textnormal{Ca}^\frac{2}{3}}.
\end{equation}

This scaling collapses the data (Fig.~\ref{fig:scaling} b) suggesting that pinch-off \emph{is} controlled by the fluid layer draining from the ball. However, if the exponent of Capillary number ($2/3$ in the above) is treated as a free parameter, the optimal exponent to scale the data is approximately $0.1$ indicating that $\textnormal{Ca}^\frac{2}{3}$ is not the underlying coating mechanism. The Weber numbers ($\textnormal{We}=\rho u^2R/\gamma$) of our experiment are in the range 1--100 so that, despite the low capillary numbers, surface tension plays a less important role in the coating mechanism than the inertia of the ball \cite{quere1999fluid}.

At these Weber numbers and low to intermediate overall Reynolds numbers, we can imagine that the coating is formed by the viscous interaction between ball and fluid i.e.~that the coating layer scales with the thickness of the viscous boundary layer entrained by the ball. This visco-intertial layer forms through a balance of the acceleration of the entrained fluid, order $\rho u^2/R$, with the viscous force that leads to that acceleration, order $\mu u/s^2$, providing
\begin{equation} \label{eqn:visc_i}
s \sim \sqrt{\frac{\nu R}{u}}.
\end{equation}
In this visco-inertial coating regime the draining Reynolds number and time scale are
\begin{equation} \label{eqn:sca_vi}
\textnormal{Re}_{dv} \sim g\sqrt{\frac{1}{\nu}\left(\frac{R}{u}\right)^3}\quad \mathrm{and}\quad T_{dv} \sim \frac{1}{g}\sqrt{\frac{\nu u}{R}},
\end{equation}
(combining Eqns.~\ref{eqn:draining} and \ref{eqn:visc_i}), leading to the scaling shown in Fig.~\ref{fig:scaling} c. These data show a clear difference in behaviour with ball size - the smallest balls (shown as stars and upward pointing triangles) having much shorter non-dimensional pinch-off times than the larger balls. At the Reynolds numbers of our experiments the visco-inertial balance of Eqn.~\ref{eqn:visc_i} may apply to the smallest ball sizes, but the coating mechanism is already becoming inviscid for the larger balls. 

In the inviscid limit we anticipate that the fluid carried with the ball is the added mass \cite{crowe1998multiphase} of the submersed ball. That is, the additional liquid mass, $m_a$, that contributes to the effective inertia of the ball as it is winched through the reservoir, where $m_a=\frac{2}{3}\pi R^3\rho$. On exiting the reservoir, this added mass therefore leads to a characteristic layer thickness that, since $m_a \sim R^3$, behaves as $s \propto R$ - depending only on the ball radius. The draining Reynolds number and timescale for an inertially coated ball scale as
\begin{equation}\label{eqn:di}
\textnormal{Re}_{di}  \sim \frac{gR^3}{\nu^2}
\quad\textnormal{and} \quad
T_{di} \sim \frac{\nu}{Rg},
\end{equation}
and this scaling shown in Fig.~\ref{fig:scaling}~d. The data collapse appears convincing, with no dependence on ball size, fluid rheology or the presence of fine sediments in the fluid. 

To summarise, we have considered various scaling models for the pinch-off time of a fluid tendril connecting a ball to the fluid from which it has been winched. This pinch-off can occur close to the reservoir or at the ball base in `surface' or `ejecta' seals respectively, depending on overall Reynolds number. We have shown that pinch-off of the tendril is controlled by Stokes flow in the thin layer of fluid viscously draining from the ball. This layer is itself formed as the ball exits the reservoir. At the intermediate overall Reynolds numbers studied, the coating layer thickness scales with the inertial added mass of the ball. Although dilute non-Newtonian suspensions lead to different flow structure and appearance, the pinch-off time, our main focus here, follows the same scaling as Newtonian fluids.

\end{document}